\documentstyle{article}

\begin{document}
\noindent Preprint  KONS-RGKU-95-2 \hfill quant-ph/9503006\\[1cm]

\begin{center}
\Large\bf
A pragmatic approach to the problem of the self-adjoint
extension of Hamilton operators with the Aharonov-Bohm potential
\\[1cm]
\normalsize\bf
J\"urgen Audretsch$^{\dagger,}$\footnote{{\it e-mail:
Juergen.Audretsch@uni-konstanz.de}},
Ulf Jasper$^{\dagger,}$\footnote{{\it e-mail: Ulf.Jasper@uni-konstanz.de}},
and Vladimir D. Skarzhinsky$^{\dagger, \ddagger,}$\footnote{{\it e-mail:
vdskarzh@sgi.lpi.msk.su}
\hfill to appear in J.~Phys.~A}
\\[0.5cm]
\normalsize
$^\dagger$Fakult\"at f\"ur Physik der Universit\"at Konstanz\\
Postfach 5560 M 674, D-78434 Konstanz, Germany \\[0.5cm]
$^\ddagger$P.~N.~Lebedev Physical Institute\\
Leninsky prospect 53, Moscow 117924, Russia
\\[1cm]

\begin{minipage}{15cm}
\begin{abstract}
\noindent We consider the problem of self-adjoint extension of Hamilton
operators for charged quantum particles in the pure Aharonov-Bohm potential
(infinitely thin solenoid). We present a pragmatic approach to the problem
based on the orthogonalization of the radial solutions for different quantum
numbers. Then we discuss a model of a scalar particle with a magnetic moment
which allows to explain why the self-adjoint extension contains arbitrary
parameters and give a physical interpretation.

\end{abstract}
\end{minipage}

\end{center}
\vspace{1cm}

\section{Introduction}

The theoretical prediction of the Aharonov-Bohm (AB) effect \cite{Aharonov59}
in 1959 was one of the most intriguing results of quantum theory. Now AB-effect
has been long recognized for its crucial role in demonstrating the specific
status of electromagnetism in quantum theory. Beside usual local influence of
electric and magnetic fields on charged particles it manifests nonlocal quantum
effects from electromagnetic fluxes $\Phi=\oint A_i d x_i$ or the corresponding
phase factors,  $\exp(ie\oint A_i d x_i)$ \cite{Yang75}. Shifting phases of
wave functions these gauge invariant factors influence interference patterns,
energy spectra of quantum particles, and cause other quantum phenomena (for a
detailed exposition of theoretical and experimental attempts to investigate
the AB-effect see \cite{Olariu85, Peshkin89}). One of these phenomena is
scattering of charged particles by a magnetic string \cite{Aharonov59} which
arises due to distinctive interference of the particle wave. It was shown in
\cite{Sereb86} that AB-scattering is accompanied by electromagnetic radiation,
and its angular distribution and polarization were calculated in \cite{Sereb86,
Gal'tsov90}. A clear example of the AB-effect for bound states is the splitting
of Landau energy terms for charged particles in a uniform magnetic field
\cite{Lewis83}. In addition there exist remarkable applications of the
AB-effect to solid state physics \cite{Lee85, Bergmann84}.

The issue of spin appended further peculiarity to the status of the AB-effect.
It was found that the interaction between the magnetic momentum of a charged
particle and  the magnetic field of the AB-string changes essentially the
behavior of the wave functions at the magnetic string \cite{Gerbert89, Hagen90,
Hagen91}. In the case of attraction this interaction increases the probability
to find the particle near the magnetic string so that an irregular component
inevitably appears in the radial solution. It becomes quite obvious that the
Hamilton operator is not self-adjoint in this case. It is the role of the
irregular solutions to which we want to draw attention.

Characteristic for the AB effect is that a magnetic field is localized inside
a solenoid and vanishing outside. There are many physical realizations for
this. But in practice physical processes, as for example quantum field
theoretical processes, can only be studied in detail when reference to a much
simpler limiting case is made: the infinitely thin and infinitely long,
straight solenoid (pure AB case). This is therefore the important situation
to be studied for different matter field equations. For the radial equations
in the Schr\"odinger and Dirac case, irregular solutions cannot be excluded
by the normalization condition. At this point usually the mathematically
cumbersome procedure of self-adjoint extension of the respective Hamiltonian
is applied \cite{Gerbert89}. It is the first aim of this paper to point out
an equivalent pragmatic approach to the problem which is quick and transparent.

The resulting self-adjointness conditions don't fix the solutions but still
contain open parameters \cite{Hagen90, Hagen91}. Their appearance reflects
the fact that different original physical situations are described by the
same pure AB case. To discuss this in detail is the second aim of this paper.
We mention that the problem of bound states for quantum particles with magnetic
moment in the AB-potential is considered in detail in papers \cite{Bordag93,
Bordag94}.

\bigskip

This paper is organized as follows. In section 2 we consider radial solutions
to wave equations in the presence of the pure AB-potential and discuss a
mathematical problem which arises due to the singular behavior of the
potential. The problem of self-adjoint extension for the Hamilton operator
to wave equations with AB-potential is discussed in section 3. We present
the direct approach to the problem based on the orthogonalization of the
radial solutions with different quantum numbers. In section 4 the problem
of physically adequate choice of the solution is discussed. Than we discuss
a model of a scalar particle with a magnetic moment which allows to illustrate
why the standard method of self-adjoint extension contains an arbitrary
parameter.

\bigskip

We use units such that $\hbar = c =1$.


\section{The pure Aharonov-Bohm case}

The {\em pure AB potential} \cite{Aharonov59} which reads in cylindrical
coordinates
\begin{equation} \label{abp}
eA_{\varphi}
= {e\Phi \over 2\pi\rho} = {\Phi \over \Phi_{0}\rho}
= {\phi\over\rho}\;,
\end{equation}
is realized by an infinitely thin solenoid lying along the
$z$-axis. The related magnetic field is localized on the $z$-axis
\begin{equation}
H_z = {\phi\over e} {\delta(\rho)\over\rho}
\end{equation}
Here $\Phi_0 = 2\pi/e$ is the flux quantum. In the
following we decompose the flux $\phi$ into an integer part $N$
and a fractional part $\delta$ with $0<\delta<1$, i.~e.~$\phi = N
+\delta$. As we will see it is the fractional part $\delta$ of the
magnetic flux  which produces all physical effects.

The corresponding stationary Schr\"odinger equation reads
\begin{equation} \label{scheq}
{1\over 2M}\left(-i \vec{\nabla} - e \vec{A} \right)^2
\psi_j (\rho, \varphi, z)
= E\ \psi_j (\rho, \varphi, z) \;,
\end{equation}
where $j$ is a collective
index for quantum numbers.
After separating the angular and $z$-dependence with the ansatz
\begin{equation}
\psi_j (\rho, \varphi, z) = e^{i p_3 z} e^{i l \varphi} R_l(\rho) \;,
\end{equation}
we find that the radial part $R_l(\rho)$ of the
solution obeys the Bessel equation
\begin{equation} \label{re}
h_l R_l
\equiv R''_l + {1\over \rho }R'_l - {(l-\phi)^2\over \rho^2}R_l
= -p_\perp^2 R_l  \;,
\end{equation}
where $2 M E = p_\perp^2+p^2_3$ and $l$ is the angular momentum
projection. The general solution of this equation,
\begin{equation} \label{rs}
R_l = a_l J_{|l-\phi|}(p \rho) + b_l J_{-|l-\phi|}(p \rho),
\end{equation}
contains regular parts with Bessel functions of positive orders as
well as irregular parts with Bessel functions of negative orders.
For those $l$ with $|l-\phi| >1$, i.~e.~for $l\neq N$ or $N+1$ the
normalization condition
\begin{equation} \label{nc}
\int_0^{\infty} R_l(p' \rho) R_l(p \rho) \rho d\rho =
{\delta(p-p')\over\sqrt{p p'}}
\end{equation}
eliminates the irregular parts which diverge at $\rho=0$. Accordingly we have
$a_l = 1$, $b_l= 0$ in this case. But for $l=N$ or $N+1$ the Bessel functions
of positive and of negative order both are square integrable and we cannot
fix the coefficients in this way so that irregular solutions are not excluded.
These modes require a separate discussion. In the following we want to
contribute to a clarification of this problem.

\bigskip

A similar situation occurs for the Dirac equation, Here it is also
possible to separate variables. One finds for the $\rho$-depending
part of each spinor component a radial equation of the type
\begin{equation} \label{Dre}
\widetilde{h}_l R_l
\equiv R''_l + {1\over \rho }R'_l - {(\nu-\phi)^2\over \rho^2 }R_l
+ s  \phi {\delta(\rho)\over\rho} R_l
= - p^2_\perp R_l  \;,
\end{equation}
where $p_{\perp} = \sqrt{p^2 - p^2_3} = \sqrt{E_p^2 - M^2 - p^2_3}$
is the radial momentum. For different two-spinor components $s$ takes the
values $\pm1$, and $\nu=l$ or $l+1$.
Note the appearance of the $\delta$-function. It arises from the
$\sigma^{\mu\nu}F_{\mu\nu}$ term which is implicitly contained in the
Dirac equation. For the pure Aharonov-Bohm potential it reduces to
$\sigma^z B_z \sim \delta(\rho)/\rho$. In the open interval $(0,\infty)$ we
find in going back to the full first order Dirac equation as solutions for
the components of the two-spinors
\begin{equation} \label{sds}
R_l^1 = a_l J_{l-\phi}(p_{\perp}\rho)+b_l J_{-l+\phi}(p_{\perp}\rho)
\end{equation}
and
\begin{equation} \label{sds2}
R_l^2 = a_l J_{l-\phi+1}(p_{\perp}\rho)
- b_l J_{-l+\phi-1}(p_{\perp}\rho)
\end{equation}
The normalization condition here is more complicated but of the same type as
(\ref{nc}). It shows that for all $l\neq N$ the Bessel functions of negative
orders must be removed. For nonnegative $N$ for example one finds $b_l=0$ for
$l>N$ and $a_l=0$ at $l<N$. Here only one critical mode occurs. For $l=N$ each
spinor component contains an irregular part. It is not possible to remove all
of them at the same time. Therefore for $l=N$ at least one component of the
two-spinors becomes irregular at $\rho = 0$. So in the Dirac case the problem
of irregular solutions of the radial equation is even more evident.

\bigskip

Although the radial equations (\ref{re}) and (\ref{Dre}) in the Schr\"odinger
and the Dirac case
are essentially the same one finds different numbers of critical
modes and different conditions for the coefficients $a$ and $b$. This
is a consequence of the definition of the respective adjoint operator
(see (\ref{sc1}) below) which depends on the scalar product of the Hilbert
space.


\section{The self-adjoint extension and a simple equivalent
procedure}

The fact that the irregular radial solutions of the
Schr\"odinger and Dirac equations cannot be ignored is related to the
fact that the respective Hamilton operators $h_l$ and $\widetilde{h}_l$
are {\em not self-adjoint}.
Self-adjointness however is needed for a unitary time evolution.

Consider the radial equation (\ref{re}) for $l=N$ or $N+1$. The domain of
the \lq radial Hamilton operator\rq\ $h_l$ is given by the set $D(h_l)
= \left\{ R_l \in {\cal L}^2((0,\infty),\rho d\rho) \; |\; R_l(0) =
0\right\}$, i.e.~the square integrable functions with support away
from the origin which have a regular limit for $\rho\to0$.

The adjoint operator $h_l^\dagger$ is constructed in the following way:
The domain $D(h_l^\dagger)$ of $h_l^\dagger$ consists of all states
$S_l$ for which there exists a state $S_l^\prime$ such that
\begin{equation} \label{sc1}
\langle R_l|h_l^\dagger S_l \rangle
= \langle R_l| S_l^\prime \rangle
\end{equation}
for all states $R_l \in D(h_l)$. Then $h_l^\dagger$ is defined by
$h_l^\dagger S_l = S_l^\prime$. It turns out that the domains of
$h_l$ and $h_l^\dagger$ are not the same. $D(h_l^\dagger)$ also
contains the irregular solutions and $h_l$ is therefore not
self-adjoint.

A detailed analysis of the operator $h_l$ shows that it is possible to extend
its domain in order to make it self-adjoint.
This extension essentially consists in the inclusion of irregular solutions in
$D(h_l)$. But because of its mathematical complexity we shall not present this
procedure here.For an accurate and mathematically exact treatment of the method
of self-adjoint extensions we refer to \cite{ReedSimon}.

For the Schr\"odinger case this scheme of {\em self-adjoint extension} leads
to the self-adjointness conditions
\begin{equation}\label{ab0}
{b_N\over a_N}
= \alpha_0 \left({p\over M}\right)^{2\delta}\;,
\end{equation}
and
\begin{equation} \label{ab1}
{b_{N+1}\over a_{N+1}}
= \alpha_1 \left({p\over M}\right)^{2(1-\delta)} \;,
\end{equation}
correlating the open parameters in (\ref{rs}) where $\alpha_0$ and $\alpha_1$
are arbitrary real numbers called extension parameters. We can
express equations
(\ref{ab0}) and (\ref{ab1}) in terms of new boundary conditions replacing
$R_l(0)=0$:
\begin{equation} \label{newbc}
\lim_{\rho\to 0} R_l(p\rho)
\propto (M\rho)^{|l-\delta|} - \tilde{\alpha}(M\rho)^{-|l-\delta|} \;,
\end{equation}
where
\begin{equation}
\tilde{\alpha}
= 2^{2|l-\delta|} {\Gamma(|l-\delta|)\over \Gamma(-|l-\delta|)}\;\alpha\;.
\end{equation}
Therefore, in order to make $h_l$ self-adjoint we have to choose as
domain the square integrable functions that satisfy the boundary
condition (\ref{newbc}) thus allowing an irregularity at $\rho=0$.

The self-adjoint extension that is constructed in this way depends on
the two parameters $\alpha_0$ and $\alpha_1$. It is a characteristic trait
of this procedure that they remain open and
cannot be determined without any additional information. Because of
the relation to boundary conditions it is obvious that they must be
connected with the physical details of the flux distribution inside the
solenoid of the underlying original model from which the pure AB case was
obtained in a limiting procedure.

\bigskip

For the Dirac case the first order Hamilton operator reads
\begin{equation}
\left(\begin{array}{cc}
s M & i\partial_\rho + i{l+1-\phi \over \rho}\\
i\partial_\rho - i{l-\phi \over \rho} & -s M \\
\end{array}\right)
\end{equation}
for eigenstates of the spin-$z$ operator $ S_3 = \gamma^0 \Sigma_3
+ \gamma^5 {P_3\over M} $ and the self-adjointness condition for $l=N$
obtained in the same involved mathematical procedure of self-adjoint extension
takes the form
\begin{equation} \label{tan}
{b_N\over a_N} =
\alpha {M\over E + sM}\left({p_{\perp}\over M}\right)^{2\delta} \;,
\end{equation}
where $\alpha$ is an arbitrary dimensionless number. For a different spin
projection e.~g.~or helicity eigenstates it differs only by a factor which
is independent on $p_\perp$.

\bigskip

We present now an alternative, {\em pragmatic approach} to the problem of
the indetermined parameters in (\ref{rs}) or (\ref{sds}) and (\ref{sds2})
respectively. It is simple and quick.

\noindent Consider again the Schr\"odinger case. Because regular and irregular
parts both are square integrable we take as solutions for $l=N$ and $N+1$
\begin{equation} \label{rs0}
R_N = a_N J_{\delta}(p \rho) + b_N J_{-\delta}(p \rho)\;,
\end{equation}
and
\begin{equation}\label{rs1}
R_{N+1} = a_{N+1} J_{1-\delta}(p \rho) + b_{N+1} J_{-1+\delta}(p \rho)\;.
\end{equation}
The observation is now that these solutions are not {\em orthogonal} for
different $p$ and $p'$
\begin{equation} \label{o}
\int_0^{\infty} R_l(p' \rho) R_l(p \rho) \rho d\rho \neq
{\delta(p-p')\over\sqrt{p p'}} \;.
\end{equation}
This results from the cross terms containing integrals over Bessel
functions of opposite orders. Using the well known formula
\begin{equation} \label{++}
I_+ =
\int_0^{\infty} \rho d\rho J_{\delta}(p\rho) J_{\delta}(p'\rho)
= {1\over \sqrt{p p'}}\delta (p-p') \end{equation}
and the for our purpose newly developed formula
\begin{equation} \label{--}
I_- =
\int_0^{\infty} \rho d\rho J_{\delta}(p\rho)
J_{-\delta}(p'\rho) = {1\over \sqrt{p p'}}\delta (p-p')\cos\pi\delta
+{2\sin\pi\delta\over \pi (p^2 - p'^2)}\left({p\over
p'}\right)^{\delta}
\end{equation}
we can calculate the integral (\ref{o}). To our knowledge the integral
(\ref{--}) has not been solved before. It can be derived from the known
indefinite integral (5.53) of \cite{Gradshteyn80} by extending the range
of integration to $(0,\infty)$ and using the asymptotic form of the Bessel
functions. The relation
$ \lim_{L\to\infty}{\sin(x L)/ x} = \pi\delta(x)$
then leads to (\ref{--}). It is easy to show that the non-$\delta$ terms
which arise from (\ref{--}) are canceled if the coefficients $a$ and $b$
fulfill (\ref{ab0}) and (\ref{ab1}). These conditions can therefore be
derived this way. The same procedure can also be applied in the Dirac case
and easily leads to (\ref{tan}).

Thus the orthonormality condition lead us directly to the
self-adjointness condition thereby circumventing the mathematically
cumbersome procedure of self-adjoint extension.
This is of course not just a coincidence but is related to the
fact that a self-adjoint operator possesses a complete set of
orthonormal eigenstates.

\bigskip

The practical relevance of the pragmatic approach described above is to be
seen in the fact that it shortens for example the calculations of quantum
electrodynamical effects outside thin solenoids. It will be used in a
subsequent discussion \cite{qedab} of the bremsstrahlung emitted by an
electron which is scattered by the external Aharonov-Bohm potential.


\section{The open parameters $\alpha$ and their physical meaning}

The pure AB case is an approximative description of a whole class of
real physical situations. All the different configurations which in the
limit of
vanishing solenoid radius and fixed flux $\Phi$ lead to the AB
potential (\ref{abp}) are described by it. The appearance of the open
extension parameters $\alpha$ in the pure AB case reflects this.
Different $\alpha$ correspond to different original situations.
Therefore we have to go back to the original situation to find
the specific values of $\alpha$.

\bigskip

All cylindrically symmetric magnetic fields
which vanish for $\rho > \rho_0$ so that $A_\varphi = {\phi\over e\rho}$
and satisfy
$$
\lim_{\rho_0\to0}\int_0^{\rho_0} H(\rho) \rho d\rho = 0
$$
lead in the AB limit $\rho_0\to0$ to the same values of $\alpha$. This was
shown by Hagen
\cite{Hagen90} for the Dirac case but applies to spinless and
nonrelativistic particles as well because the radial equations are
identical.

\noindent For Schr\"odinger particles we have
\begin{equation} \label{speschr}
\alpha_0= 0 \;,\quad \alpha_{1}= 0\;,
\end{equation}
and for Dirac particles, depending on the mutual interaction of spin
and magnetic field,
\begin{equation} \label{spedir}
\alpha = \left\{\begin{array}{ll}
0 &{\rm for }\, s\phi < 0 \\
\infty &{\rm for }\, s\phi > 0 \;. \\
\end{array}\right.
\end{equation}
The Dirac particle carries a magnetic moment $\mu = (e/2 M) s$ ($s=\pm 1$)
which interacts with the magnetic field resulting in a potential energy
$-\vec{\mu}\vec{H}$. Therefore it suffers an attractive force if $s\phi>0$,
i.~e.~if magnetic moment and magnetic field are parallel
($\vec{\mu}\vec{H}>0$), which
leads to an enhancement of the wavefunction. For the Schr\"odinger particle
such an interaction is not present and thus the wavefunction always stays
regular at $\rho=0$.

\bigskip

Because the pure AB case allows parameter values different from
(\ref{speschr}) and (\ref{spedir}) it is more general. It describes also
physical situations different from the one sketched above.
\footnote{We thank Dr. Michael Bordag
for a fruitful discussions about this problem.} What is therefore the
physical meaning of the {\em nontrivial parameters} $0<\alpha<\infty$?
We will give an example.

We saw that the $\vec{\mu}$-$\vec{H}$ interaction is responsible for the
enhancement of the Dirac wave function near $\rho=0$. Therefore we will
consider the influence of an additional interaction of this type for a
Schr\"odinger particle thus modifying the Schr\"odinger theory. If, in
the limit of vanishing solenoid radius, this new model gives the same
radial equations as before the self-adjoint extension procedure will apply
here too. This indeed is the case because the additional interaction is in
this limit localized to $\rho=0$ and the radial equation remains unchanged
at $\rho>0$.
Now, in order to fix $\alpha$ we have to return again to the original
physical situation and study the limit of vanishing solenoid radius.We
will show that in deed it can lead to nonzero $\alpha$ values.

Let us consider a situation in which the magnetic flux is located on the
surface of a
cylinder. Then the vector potential and magnetic field are given by
\begin{equation} \label{Hag1}
eA_{\varphi} = {\phi\over\rho}\;\Theta(\rho-\rho_0)\;,
\quad eH_z = {\phi\over\rho_0}\;\delta(\rho-\rho_0)\;.
\end{equation}
We modify the Schr\"odinger equation in assuming that the particle
carries a magnetic moment $\vec{\mu}$ that couples to the magnetic
field $\vec{H}$:
\begin{equation} \label{mHscheq}
\left[{1\over 2M}
\left(-i \vec{\nabla} - e \vec{A} \right)^2 - \vec{\mu} \vec{H}\right]
\psi_j (\rho, \varphi, z)
= E\ \psi_j (\rho, \varphi, z)\;.
\end{equation}
We put $\mu_z = g {e\over 2M}$ but do not specify $g$ and find for
the radial equation
\begin{equation} \label{Hre}
R''_l + {1\over \rho }R'_l - {(l-\phi)^2\over \rho^2}R_l
+ {g\phi\over\rho_0}\;\delta (\rho-\rho_0)R_l + \epsilon R_l = 0 \;,
\end{equation}
where $\epsilon = 2ME$.

The interior and exterior solutions of the
radial equation (\ref{Hre}) are given by
\begin{equation}
R_l = \left\{
\begin{array}{ll}
\label{Hrsi}
c_l \; J_{|l|}(p\rho)\;, & {\rm for}\; \rho<\rho_0 \\
\label{Hrse}
a_l \; J_{|l-\phi|}(p\rho) + b_l \; J_{-|l-\phi|}(p\rho)
\;, &{\rm for}\; \rho >\rho_0
\end{array}\right.
\end{equation}
and the matching conditions read
$$
R_l^{int}(\rho_0) = R_l^{ext}(\rho_0) \;, \quad
\rho R_l^{int^\prime}(\rho_0)
= \rho R_l^{ext^\prime}(\rho_0) + g \phi R_l(\rho_0) \;.
$$
They lead to
\begin{equation} \label{bda}
{b_l \over a_l} =
-{ J_{|l-\phi|}^\prime(p\rho_0) J_{|l|}(p\rho_0)
-J_{|l-\phi|}(p\rho_0)[J_{|l|}^\prime(p\rho_0)
-{g\phi\over p\rho_0}J_{|l|}(p\rho_0)]
\over
J_{-|l-\phi|}^\prime(p\rho_0) J_{|l|}(p\rho_0)
-J_{-|l-\phi|}(p\rho_0)[J_{|l|}^\prime(p\rho_0)
-{g\phi\over p\rho_0}J_{|l|}(p\rho_0)]} \;,
\end{equation}
which fixes $a_l$ and $b_l$ for arbitrary $\rho_0$.

Inserting the series representation of the Bessel function, (\ref{bda})
becomes in the limit of vanishing radius $\rho_0\to 0$
\begin{equation} \label{mHHab}
{b_l \over a_l} \to
{|l-\phi| - |l| +g\phi\over |l-\phi| + |l|  -g\phi} \cdot
{\Gamma(-|l-\phi|) \over \Gamma(|l-\phi|)}
\left({p\rho_0\over 2}\right)^{2|l-\phi|} \;.
\end{equation}
We see that for $\rho_0=0$ we have again $b_l=0$ for all $l$ unless the
denominator in
(\ref{mHHab}) becomes zero,
\begin{equation}
|l-\phi| + |l| - g\phi = 0 \;.
\end{equation}
For this case we must consider the next term of the series in eq.~(\ref{bda})
and find that this can only happen for $l=N$ or $N+1$.

The particular physical situation (\ref{Hag1}) treated by the modified
Schr\"odinger equation (\ref{mHscheq}) can also approximately
(limit $\rho_0\to0$) be represented as a particular pure AB case, if the
self-adjointness conditions (\ref{ab0}) and (\ref{ab1}) are fulfilled.
Comparison with (\ref{mHHab}) shows that this is indeed the case if $g$
satisfies for $l=N$ the condition
\begin{equation} \label{g0}
g_N = {1\over N+\delta} \cdot
{\Gamma(-\delta)
\left({M\rho_0\over 2}\right)^{2\delta}(|N|-\delta)
+ \alpha_0 \Gamma(\delta)(|N|+\delta)
\over
\Gamma(-\delta)
\left({M\rho_0\over 2}\right)^{2\delta}+ \alpha_0 \Gamma(\delta)}
\;,
\end{equation}
and for $l=N+1$:
\begin{equation} \label{g1}
g_{N+1} = {1\over N+\delta} \cdot
{\Gamma(-1+\delta)
\left({M\rho_0\over 2}\right)^{2(1-\delta)}(|N+1|-1+\delta)
+ \alpha_1 \Gamma(1-\delta)(|N+1|+1-\delta)
\over
\Gamma(-1+\delta)
\left({M\rho_0\over 2}\right)^{2(1-\delta)}+ \alpha_1
\Gamma(1-\delta)} \;,
\end{equation}
Thus we see that the pure AB case may also describe the \lq modified\rq\
Schr\"odinger particle that suffers an additional
$\vec{\mu}\vec{H}$-interaction if its $g$-factor has the properties
(\ref{g0}) and (\ref{g1}). The extension parameters $\alpha_0$ and
$\alpha_1$ are then determined by $g_N$ and $g_{N+1}$ and need not to be
zero as it is the case of the Schr\"odinger equation (\ref{scheq}).

\bigskip

In general the conditions (\ref{g0}) and (\ref{g1}) are rather exotic
because the $g$-factor depends on the angular momentum $l$ and on the
solenoid parameters $\rho_0$ and $\Phi$ (more precisely on $\delta$).
The reason is that we are still dealing with the whole range
$0<\alpha<\infty$ of nontrivial parameters, i.~e.~with all the parameters
which do not fulfill (\ref{speschr}). Particular combinations of nontrivial
parameters, and this is enough for our purpose, can be combined with reasonable
physical situations:
For $N\geq 0$ we may for example choose the nontrivial combination
$\alpha_0\neq0$ and $\alpha_1=0$. Then we have $g_{N+1}=1$ and $g_N$ still
depends on $\rho_0$ and $\phi$ but it approaches the same value $+1$ in the
pure AB case as $\rho$ goes to zero,
\begin{equation} \label{g+}
g_N \to 1 + {1\over \alpha_0}{N-\delta\over N+\delta}
{\Gamma(-\delta)\over \Gamma(\delta)}
\left( {M\rho_0\over 2}\right)^{2\delta} \;.
\end{equation}
For $N\leq0$ we may choose $\alpha_0 = 0$, $\alpha_1 \neq 0$ and find
$g_N=-1$ and the same value for $g_{N+1}$
\begin{equation} \label{g-}
g_{N+1} \to -1 -
{1\over \alpha_1}{N + 2-\delta\over N+\delta}
{\Gamma(-1+\delta)\over \Gamma(1-\delta)}
\left( {M\rho_0\over 2}\right)^{2(1-\delta)}
\end{equation}
in the limit of vanishing solenoid radius $\rho_0$.
Thus we have obtained the result that in the pure AB case the self-adjoint
extension in which one of the parameters is nonzero may describe a
\lq modified\rq\ Schr\"odinger particle obeying (\ref{mHscheq}). It
carries a magnetic moment oriented such that $\vec{\mu}\vec{H}>0$.
This model provides one possible physical
explanation of nontrivial parameter values which arise from the
self-adjoint extension method.
The $\vec{\mu}\vec{H}$-interaction that we put in by hand here is
already present in the Dirac and Pauli equation. Therefore the
inclusion of an additional (anomalous) magnetic moment can explain
particular parameter values there. This completes our presentation of a
physical situation which may be described by the pure AB case with a
nontrivial combination of extension parameters.


\section{Conclusion}

We analyzed the problem of the self-adjoint extension for the
Hamilton operator containing the pure AB potential. Using a
pragmatic approach based on a direct procedure which allows to make
radial solutions orthogonal at different quantum numbers, we reproduced in
a straightforward manner
results which follow from the standard method of self-adjoint
extension. Regression to the original physical problem leads to
definite values for the extension parameter depending on the
specific form of the interaction with the magnetic field. In the
framework of a simple model of a charged particle with an exotic
magnetic moment (modified Schr\"odinger theory) we explained why the standard
extension method
contains an arbitrary parameter and gave a physical meaning to
nontrivial values of this parameter.


\section{Acknowledgments}

V.~S.~thanks J.~Audretsch and the members of his group at
the University of Konstanz for hospitality, collaboration and many
fruitful discussions. This work was supported by the Deutsche
Forschungsgemeinschaft.


\end{document}